\documentclass[11pt]{article}
\usepackage{moriond,epsfig}

\def\Journal#1#2#3#4{{#1} {\bf #2}, #3 (#4)}

\def\PRL{\em Phys.~Rev.~Lett.\ }
\def\PRD{{\em Phys.~Rev.\ }~D}

\def\be{\begin{equation}}
\def\ee{\end{equation}}
\def\bea{\begin{eqnarray}}
\def\eea{\end{eqnarray}}

\newcommand{\mr}{\mathrm}

\newcommand{\non}{\nonumber}

\begin{document}
\vspace*{4cm}
\title{STRONGLY INTERACTING GAUGE BOSON SYSTEMS AT THE LHC 
\footnote{
presented by B.\ J\"ager at ``Recontres de Moriond -- QCD and High Energy Interactions 2009''
}
}

\author{C.\ ENGLERT$^1$, B.\ J\"AGER$^2$, M.\ WOREK$^3$, AND D.\ ZEPPENFELD$^1$}

\address{
$^1$Institut f\"ur Theoretische Physik, 
Universit\"at Karlsruhe, KIT, 76128 Karlsruhe, Germany
\\
$^2$Institut f\"ur Theoretische Physik und Astrophysik, Universit\"at W\"urzburg, \\
97074~W\"urzburg, Germany
\\
$^3$Fachbereich C Physik, Bergische Universit\"at Wuppertal, 
42097 Wuppertal, Germany
}

\maketitle\abstracts{
We explore the potential of the CERN-LHC to access strongly interacting gauge boson systems via weak-boson scattering processes with $W^+W^-jj$, $ZZjj$, and $W^\pm Zjj$ final states, focusing on the leptonic decay modes of the gauge bosons. Cross sections and kinematic distributions for two representative scenarios of strong interactions in the weak sector and all relevant background processes are computed with fully-flexible parton-level Monte-Carlo programs that allow for the implementation of dedicated selection cuts. We find that models with new resonances give rise to very distinctive distributions of the decay leptons. The perturbative treatment of the signal processes is under excellent control. 
}


\section{Introduction}
Weak gauge boson scattering reactions provide particularly promising means for gaining insight into the mechanism of electroweak symmetry breaking. At hadron colliders such as the CERN LHC this class of processes can be probed in vector-boson fusion (VBF) reactions, where the quarks emerging from the scattering protons emit weak bosons which in turn scatter off each other. The decay products of the gauge bosons emerge almost back-to-back at central rapidities, while the quarks give rise to energetic jets of relatively low transverse momenta in the forward and backward regions of the detector. Due to the lack of color exchange between the two separate quark lines, additional jet emission from the central region is strongly suppressed. 

These distinctive features allow for a separation of VBF reactions from QCD processes with {\em a priori} much larger cross sections. In Ref.~\cite{vvstrong} we have developed dedicated selection cuts which help to minimize backgrounds with respect to the signal of two ``prototype'' scenarios of strongly interacting gauge boson systems. We considered modifications of the Standard Model (SM) with a heavy and broad Higgs boson as example of a model with strong interactions in the gauge boson sector, unitarized by a scalar resonance. 
As a model with vector resonances, we adapted a Warped Higgsless scenario with an infinite tower of spin-one Kaluza-Klein (KK) excitations, $W^\pm_k$ and $Z_k$. They result from the compactification of a bulk-gauged Randall-Sundrum scenario~\cite{rs,hless} defined on a slice of a five-dimensional Anti-de Sitter space with the fifth coordinate $y$ being constrained to the interval $R\leq y\leq R'$. 
The lowest-lying states $Z_0$, $Z_1$, and $W^\pm_1$ are identified with the photon and the $Z$ and $W^\pm$ bosons of the SM. Details of the models we have used and references to the relevant literature can be found in~\cite{vvstrong,kknlo}. 

In this proceedings contribution we briefly recollect the major results of our previous work for these two representative scenarios of strongly interacting gauge boson systems at next-to-leading order (NLO) QCD accuracy.


\section{Framework of the Analysis}
Strong interactions among weak gauge bosons give rise to enhanced cross sections for longitudinally polarized gauge bosons at large invariant masses, while the scattering of the transverse modes remains perturbative. We thus define the ``signal'' of any model of strong interactions in the weak sector by the enhancement of the corresponding VBF $pp\to VVjj$ cross section over the SM prediction with a light Higgs boson. 
Throughout, $V$ stands for a $W^\pm$ or $Z$ boson. 

To clearly identify gauge boson scattering events in a hadron collider environment, the suppression of large QCD backgrounds is essential. We explore the impact of QCD $VVjj$ production, where a colored parton rather than a weak boson is exchanged between the scattering quarks and gluons, and, in the case of $W^+W^-jj$ final states, furthermore the  $t\bar t$,  $t\bar t+1\mr{jet}$, and  $t\bar t+2\mr{jets}$ production processes. For all signal and background reactions, leptonic decays of the gauge bosons are fully taken into account, retaining the spin and color correlation of all final-state particles without any approximations. Throughout, we sum over charged leptons of the first two generations and over three neutrino generations, but neglect identical-lepton interference effects. For brevity, we will refer to these reactions as $W^+W^-jj$, $ZZjj\to 4\ell jj$, $ZZjj\to 2\ell 2\nu jj$, and $W^\pm Zjj$ production, even though we are always considering leptonic final states. 
For the computation of the signal and background processes, respectively, we employ the parton-level Monte Carlo programs {\tt Vbfnlo}~\cite{vbfnlo} and {\tt Helac}~\cite{helac}.

For our numerical studies we use the CTEQ6M parton distributions  
with $\alpha_s(M_Z)=0.118$ at NLO and the CTEQ6L1 set at LO. As electroweak input parameters we choose the masses of the SM gauge bosons, $M_Z= 91.188$~GeV, $M_W=80.423$~GeV, and the Fermi constant, $G_F=1.166×\times 10^{-5}$~GeV$^{-2}$, from which we obtain $\alpha_{QED}$ and $\sin^2\theta_W$ via LO electroweak relations. Jets are recombined via the $k_T$ algorithm. 
To optimize the number of surviving signal events with respect to backgrounds, the selection cuts of Ref.~\cite{vvstrong} are applied: All jets need to be located within the detector and be well-separated from each other, 
\begin{equation}
\label{eq:cuts1}
|\eta_{j}| < 4.5\,,\quad
\Delta R_{jj}  > 0.7\,,
\end{equation}
where $\eta_j$ denotes the jet rapidity and $\Delta R_{jj}$ the separation of any pair of jets in the rapidity-azimuthal angle plane. 
The two jets of largest transverse momentum are referred to as ``tagging jets'' with
\begin{equation}
\label{eq:cuts2}
p_{Tj}^\mr{tag} > 30 ~\textnormal{GeV}\,,\quad
\Delta \eta_{jj}
     =|\eta_{j_1}^{tag}-\eta_{j_2}^{tag}| > 4\,,\quad
\eta_{j_1}^{tag} \times \eta_{j_2}^{tag} < 0\,,
\quad 
m_{jj} > m_{jj}^{min}\,,
\end{equation}
where the minimum invariant mass is given by $m_{jj}^{min} = 1000$~GeV for the $W^+W^-jj$ mode and $m_{jj}^{min} = 500$~GeV for all other channels. 
For the charged leptons, we impose 
\begin{equation}
\label{eq:cuts3}
p_{T\ell} > 20 ~\textnormal{GeV}\,,\quad
~~|\eta_\ell| < 2.5\,,\quad
m_{\ell\ell} > 15 ~\textnormal{GeV}\,,\quad 
\Delta R_{\ell j} > 0.4\,,\quad
\eta_{j,min}^{tag} < \eta_\ell < \eta_{j,max}^{tag} \,,
\end{equation}
where $m_{\ell\ell}$ is the invariant mass of 
two charged leptons of the same flavor and $\Delta R_{\ell j}$ stands for the separation of a charged
lepton from any jet. For $b$ quarks, we moreover require
$\Delta R_{\ell b} > 0.4$, even if the $b$ quark is too soft to qualify
as a jet.

In addition to Eqs.~(\ref{eq:cuts1})--(\ref{eq:cuts3}), process-specific cuts are imposed on the decay leptons. For the $ZZ\to 4\ell\,jj$ final state, we require
\be
m_{ZZ} >  500 ~\textnormal{GeV}\,,
\quad
p_T(\ell\ell) > 0.2 \times m_{ZZ}\,,
\ee
where $m_{ZZ}$ is the invariant mass of the four-lepton system, 
and $p_T(\ell\ell)$ the
transverse momentum of two same-flavor charged leptons. 
In the $ZZ jj \rightarrow 2\ell 2\nu\,jj$ decay mode, we impose 
\be
m_T(ZZ) > 500 ~\textnormal{GeV}\,,
\quad
p_{T}^{miss} > 200 ~\textnormal{GeV}\,,
\ee
with $p_{T}^{miss}$ being the transverse momentum of the neutrino system and 
\be
m^{2}_{T}(ZZ)=[ \sqrt{m_{Z}^{2}+p_{T}^{2}(\ell\ell)} + 
\sqrt{m_{Z}^{2}+(p_{T}^{miss})^2} ]^2-
[ \vec{p}_{T}(\ell\ell) +\vec{p}_{T}^{\;miss}]^2\,.
\ee
For the $W^\pm Z jj$ channel, we demand 
\be
m_T(WZ) > 500 ~\textnormal{GeV}\,,
\quad
p_T^{miss} > 30 ~\textnormal{GeV}\,,
\ee
where 
\be
m^{2}_{T}(WZ)=[ \sqrt{m^{2}(\ell\ell\ell)+p_{T}^{2}(\ell\ell\ell)} +
  |p_{T}^{miss}| ]^2 - [ \vec{p}_{T}(\ell\ell\ell) +\vec{p}_{T}^{\;miss} ]^2\, ,
\ee
with $m(\ell\ell\ell)$ and $p_{T}(\ell\ell\ell)$ denoting the invariant mass and
transverse momentum of the charged-lepton system, respectively. 

For the $W^+W^- jj$ mode, powerful cuts are needed to tame the overwhelming $t\bar t +\mr{jets}$ backgrounds:
\bea
\label{eq:ww_lep_cuts}
&p_{T\ell}>100 ~\textnormal{GeV} \,,
\quad
\Delta p_{T}(\ell\ell)=
|\vec{p}_{T,\ell_1}-\vec{p}_{T,\ell_2}|>250~\textnormal{GeV} \,,&
\non\\
&m_{\ell\ell} > 200 ~\textnormal{GeV} \,,
\quad
\min\,(m_{\ell j}) > 180 ~\textnormal{GeV} \,,&
\eea
where $\Delta p_{T}(\ell\ell)$ is the difference between the transverse momenta
of the two charged decay leptons, and $\min(m_{\ell j})$ the minimum invariant 
mass of a  tagging jet and any charged lepton. 
Moreover, for the top-quark induced backgrounds any additional jet activity in the central-rapidity region of the detector is vetoed by discarding all events with 
\be
p_{Tj}^{veto} > 25~\mr{GeV}\,,\quad
\eta_{j,min}^{tag} < \eta_j^{veto} < \eta_{j,max}^{tag}\,.
\ee
Events with jets which are positively identified as arising from $b$ quarks are also vetoed, with efficiencies depending on their transverse momenta and rapidities, cf.\ Tab.~2 in Ref.~\cite{vvstrong}.


\section{Results and Discussion}
\begin{figure}
\begin{center}
\includegraphics[height=8cm]{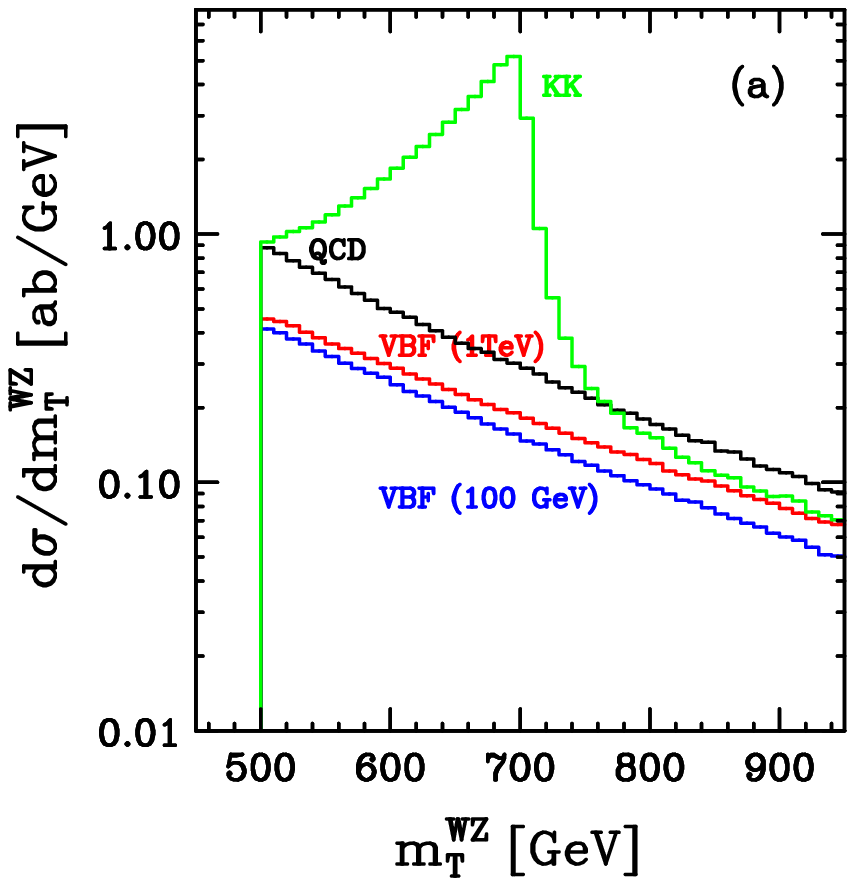}
\includegraphics[height=8cm]{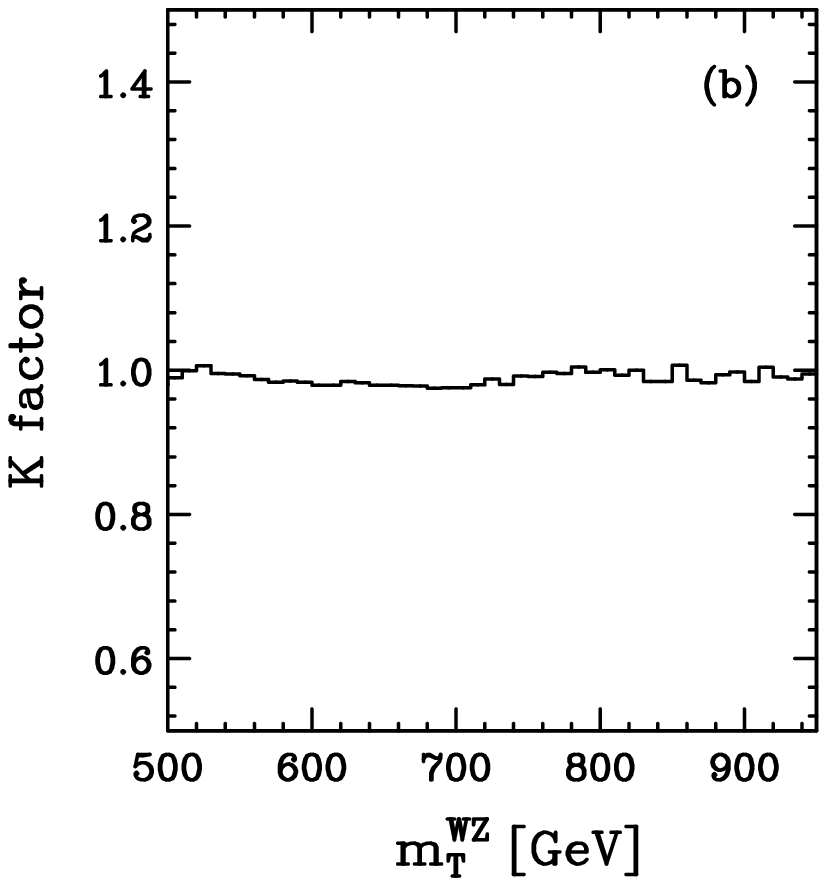}
\end{center}
\vskip -0.4cm
\caption{}{\label{fig:wpz1}
\it
Transverse mass distribution of the $W^{+}Z$ system 
for the $pp \rightarrow W^{+}Z jj $
process after imposing all dedicated selection cuts for various signal and background production processes at LO (a) and associated K factor for the Kaluza-Klein scenario (b).}
\vskip 0.4cm
\end{figure}

Cross sections for the Higgsless Kaluza-Klein scenario with $R=9.75\times 10^{-9}$~GeV$^{-1}$, the heavy Higgs scenario, and the sum of all backgrounds are calculated with the selection cuts listed above. 
A pronounced sensitivity to the extra-dimensional model occurs in the $W^\pm Z jj$ and the $W^+W^- jj$ modes, due to resonant contributions from the $W^\pm_{k=2}$ and the $Z_{k=2,3}$ excitations, respectively.  
The impact of the first Kaluza-Klein excitation with a mass of 700~GeV on the transverse mass distribution of the $W^+Z$ system in the $W^+Zjj$ mode is illustrated by Fig.~\ref{fig:wpz1}~(a),  
%
%
%
which shows the Warped Higgsless signal together with the relevant backgrounds. Also displayed is the distribution for the heavy Higgs scenario, which does not exhibit a resonance behavior in the $W^+ Zjj$ mode. 
The $ZZjj$ channels, on the other hand, are very sensitive to heavy scalar resonances, while they provide important control samples for the 
study of iso-vector spin-one excitations, cf.\ Ref.~\cite{vvstrong}. 

These features are very generic and do not suffer from large uncertainties due to radiative corrections for the signal cross sections and distributions. Similar to related VBF processes within the SM~\cite{vvjjnlo}, weak boson scattering reactions within the heavy Higgs and the Kaluza-Klein scenario we considered are under excellent control perturbatively. NLO-QCD corrections to total cross sections within the selection cuts introduced above are below 10~\% and affect the shapes of characteristic jet- and lepton distributions only marginally. This is exemplified by the ratio of the NLO to the LO prediction, called K factor, for the transverse mass of the $W^{+}Z$ system in Fig.~\ref{fig:wpz1}~(b), which is close to one and flat over the entire range of $m_T^{WZ}$ considered. 

In summary, signatures of strongly interacting gauge boson systems should be observable at the LHC on top of {\em a priori} large backgrounds after the application of dedicated selection cuts. 
The combined analysis of the $W^+W^- jj$, $W^\pm Z jj$, and $ZZ jj$ modes will allow to differentiate between various scenarios of new interactions in the gauge boson sector and thus help to gain insight into the mechanism of 
electroweak symmetry breaking.



\section*{Acknowledgements}
B.~J.\ would like to thank the organizers of ``Recontres de Moriond'' for the inspiring atmosphere during the conference and for financial support. 
The work presented here was funded by the Initiative and Networking Fund of the
Helmholtz Association, contract HA-101 ("Physics at the Terascale").



\end{document}